\documentclass[a4paper]{jpconf}
\usepackage{graphicx}
\newcommand {\la} {\langle}
\newcommand {\ra} {\rangle}
\newcommand {\beq} {\begin{eqnarray}}
\newcommand {\eeqn} [1] {\label{#1} \end{eqnarray}}
\begin{document}
\title{Single-particle motion at large distances in 2N+core cluster
systems near the drip line: a challenge for nuclear theory and
experiment}

\author{N.K. Timofeyuk$^{1}$, I.J. Thompson$^{2}$ and J.A.
Tostevin$^{1}$}

\address{$^{1}$ Department of Physics, University of Surrey, Guildford GU2 7XH, UK \\
$^{2}$  LLNL, PO Box 808, Livermore, CA 94551, USA\\}

\ead{N.Timofeyuk@surrey.ac.uk}

\begin{abstract}
There exists a class of nuclei
that are obtained by adding one nucleon to a 
loosely-bound nucleon-core system, for example $^{12}$Be, $^9$C, $^{18}$Ne. 
For such nuclei,  one-nucleon overlap integrals that
represent single-particle motion can strongly differ from the
standard ones due to the correlations between the two nucleons above the
core.  The possible non-standard  overlap behaviour
should be included in the interpretation of the experimental data 
derived from one nucleon removal reactions such as knockout, transfer and 
breakup, as well as the predictions of low-energy
nucleon capture  that leads  to these nuclei. We investigate
the non-standard behaviour within a three-body model and 
discuss the challenges associated with this problem.
\end{abstract}

\section{Introduction}

One of the most productive and perspective
ideas in nuclear many-body theory is
the concept of   independent single-particle motion
in  the mutually generated mean 
field. It  explains  many nuclear observables
qualitatively and provides a basis for quantative descriptions
of nuclei  within various versions of the shell model.
Over the last decade, the experimental study of single-particle
motion has shifted towards nuclei near the edge of stability 
with the emphasis on the occupation probabilities of single-particle
orbits   as indicators of the 
evolution of nuclear shell structure near the drip
line. These occupancies are usually determined by
comparing the measured cross sections of
one-nucleon  transfer, breakup, knockout
and Coulomb dissociation reactions, to those predicted theoretically.
In these calculations, the
single particle motion is represented by the overlap integral
between the  many-body
wave functions of   neighbouring  mass $A$ and $A-1$ nuclei.
Most theoretical analyses assume these overlaps can be calculated
using potential models that employ standard Woods-Saxon potentials.
With these potentials, at  radii
outside the binding interaction,
the overlap integrals achieve their asymptotic form given by  an
exponential decrease with a decay constant 
$\kappa  = \sqrt{2\mu  S_{1N}(A)}/\hbar$ determined by the nucleon 
separation energy $S_{1N}(A)$, with
$\mu$ is the $(A-1)$+$N$ reduced mass. This region gives
important contribution to the amplitudes of one-nucleon removal reactions
and is crucial to predict direct nucleon capture reactions 
relevant to nuclear astrophysics.

It has been shown recently in Ref.  \cite{Tim03}
that there exist
a class of nuclei for which one-nucleon overlaps can reach their asymptotics 
much later than usual. This class includes the nuclei 
at the limits of nuclear stability  with very similar one- and 
two-nucleon separation energies,
$S_{1N}(A)$ and $S_{2N}(A)$,   both of which are significantly smaller
than for stable nuclei. 
Such nuclei, for example $^{12}$Be, $^9$C, $^{18}$Ne,
can be considered as cluster systems of a core and two
valence nucleons. The correlations of the two nucleons outside 
of the nuclear core can lead to non-standard behaviour of
the overlap because even when one nucleon is removed
far from the center-of-mass of the  residual mass $A-1$ nucleus, it 
will still be affected by its interaction with the remaining
loosely-bound nucleon in $A-1$.

The possibility for  such an effect follows 
from the Feynman diagram approach in which the asymptotics
of the overlap integrals for many-body systems are represented by a sum of
different  Feynman diagrams. 
It has been noticed by Blokhintsev \cite{Blo01} that, although
in most cases the standard point diagram 
gives the exponential decrease with the smallest decay constant
$\kappa$, 
sometimes other diagrams can give a much slower decrease  
than that governed by the differences in initial and final binding
energies.
The Feynman diagram technique, applied in \cite{Tim03} to one-neutron 
overlaps for
 core+2N cluster systems, suggests that although the  
standard term, $\exp(-\kappa r)/r$, dominates at very large $r$, 
the contribution $\exp(-\kappa_1 r)/r^{7/2}$ from
the generalised triangle diagram, corresponding
to the $A \rightarrow  (A-2)+N+N  \rightarrow  (A-1)+N$ virtual process,
is determined by
the decay constant $\kappa_1$ that can be larger than $\kappa$ 
by only 40$\%$. It means that if the strength of the generalised
triangle diagram
is large  then its contribution can
be noticeable for range of $r$ where the
standard behaviour $\exp(-\kappa r)/r$
is traditionally assumed to be achieved.

The abnormal asymptotic behaviour was modelled in \cite{Tim03} 
for the $\la^{11}$Be$|^{12}$Be$\ra$ overlap  by 
adding a long-range potential to a local potential model. 
The resulting overlap 
was then used
to predict longitudinal momentum distributions for one-neutron 
knockout from $^{12}$Be.
Comparison of the latter to available
experimental data suggests such abnormalities 
would take place for  $n-^{11}$Be distances  between 5 and 10 fm. 
However, the contribution from the generalised triangle diagram 
can be calculated reliably only for
$r \gg |\kappa_1 - \kappa|^{-1}$, which for $^{12}$Be is   $r \gg 2.5$  fm, 
therefore, to understand
the overlap behaviour at $5 \leq r \leq 10 $ fm 
exact dynamical three-body calculations should be carried out.

Here, we calculate one-nucleon overlaps using a three-body core+2N model
and compare them to the predictions of the standard potential model 
 (Sec. 2). We discuss   effects
of non-standard overlap behaviour on nuclear reactions (Sec. 3) and
  challenge they pose for both theory and experiment (Sec.4).
 
\begin{figure}[h]
\begin{minipage}{15pc}
\includegraphics[width=13pc]{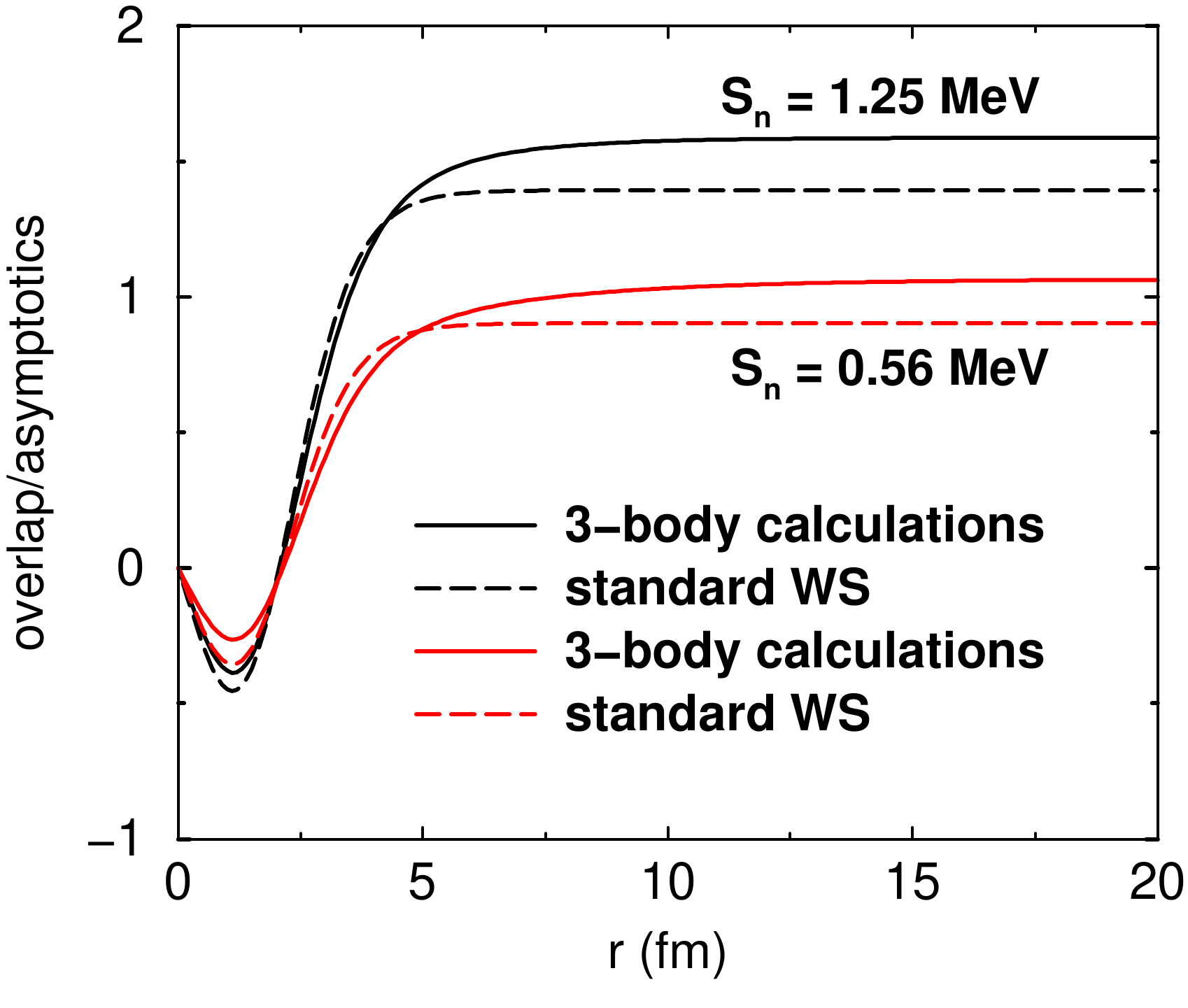}
\includegraphics[width=13pc]{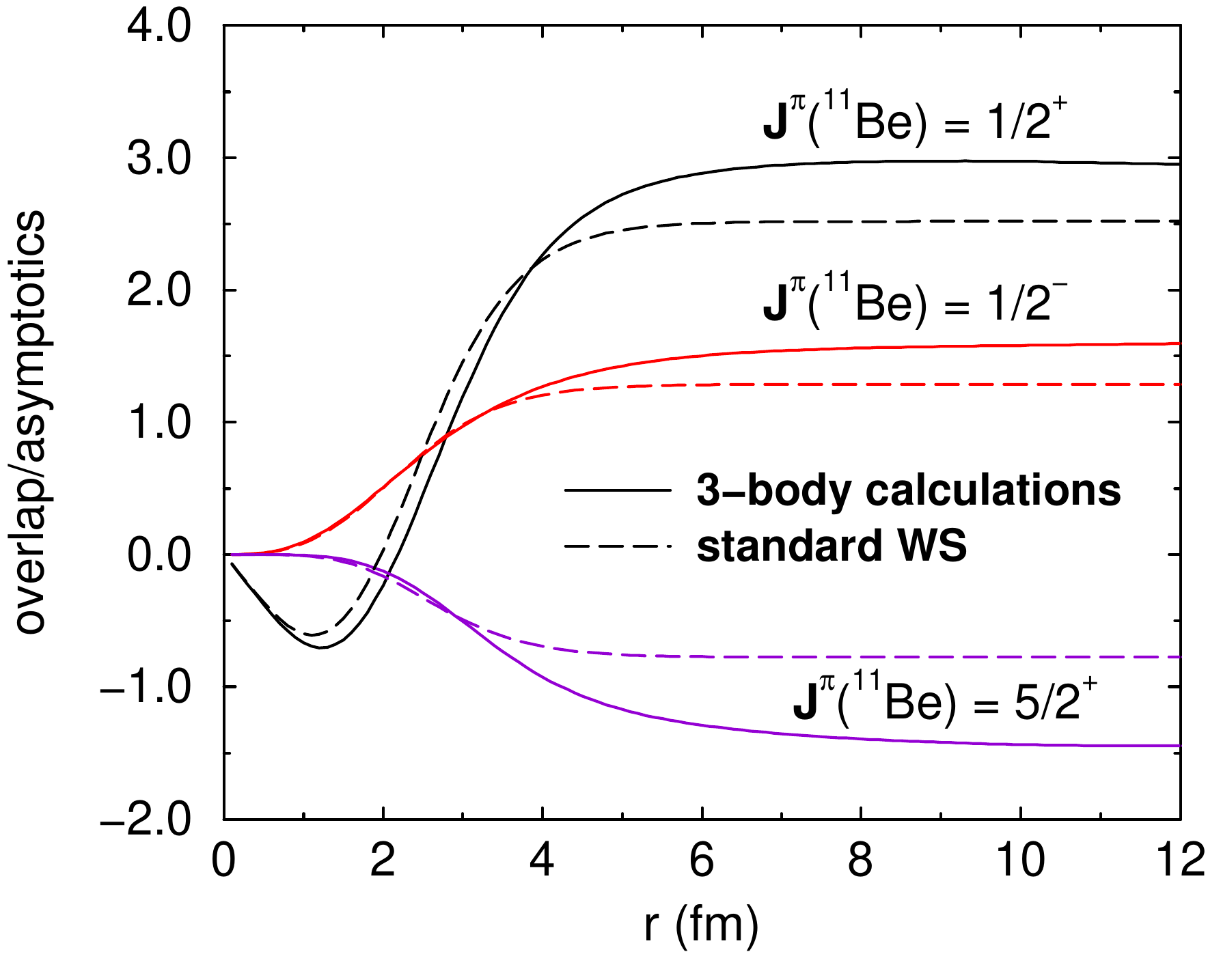}
\caption{\label{label} The $\la^{11}$Be$|^{12}$Be$\ra$ overlaps
for pseudo ($a$) and real ($b$) $^{12}$Be.}
\end{minipage}\hspace{5pc}%
\begin{minipage}{18pc}
\includegraphics[width=13pc]{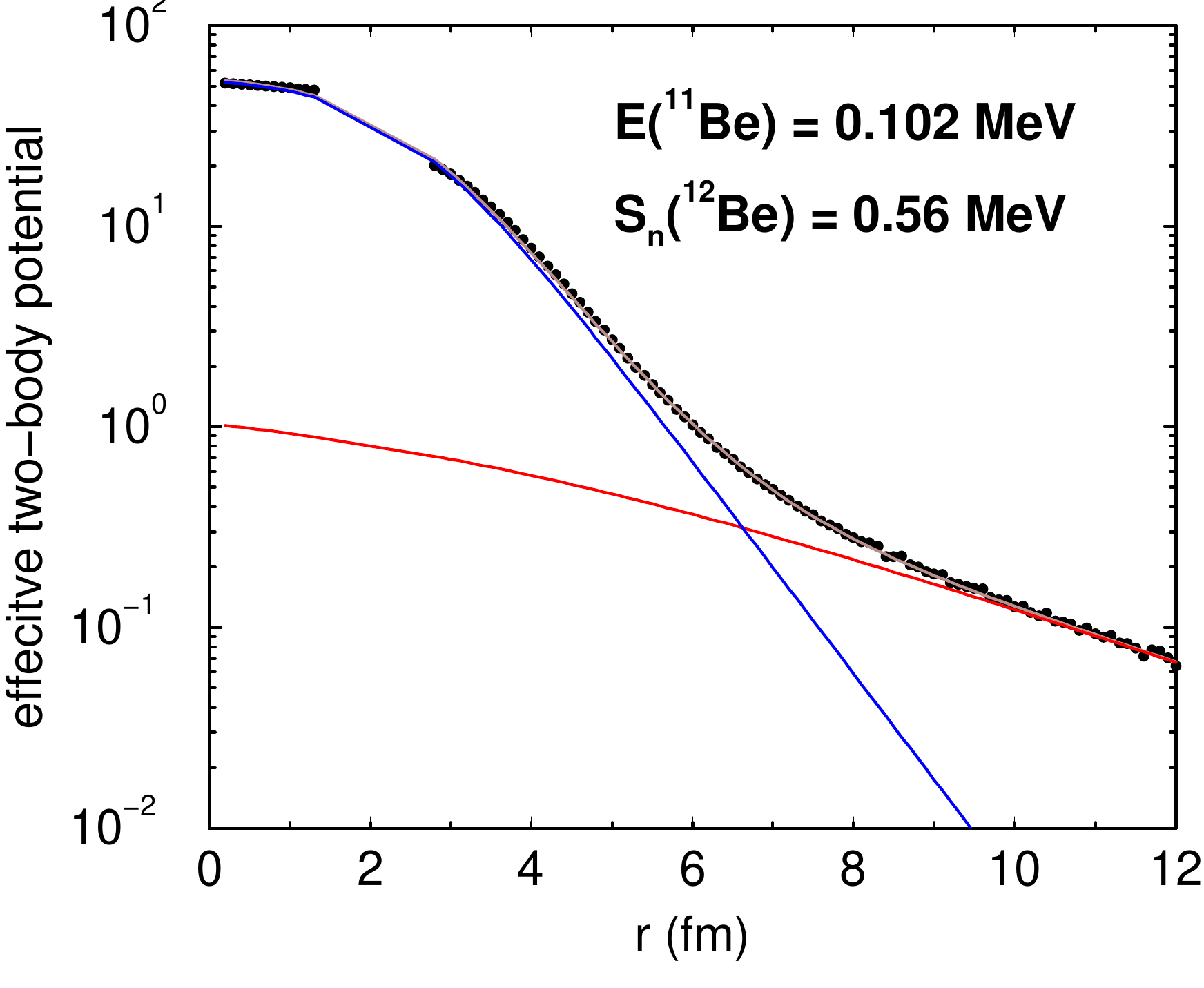}
\includegraphics[width=13pc]{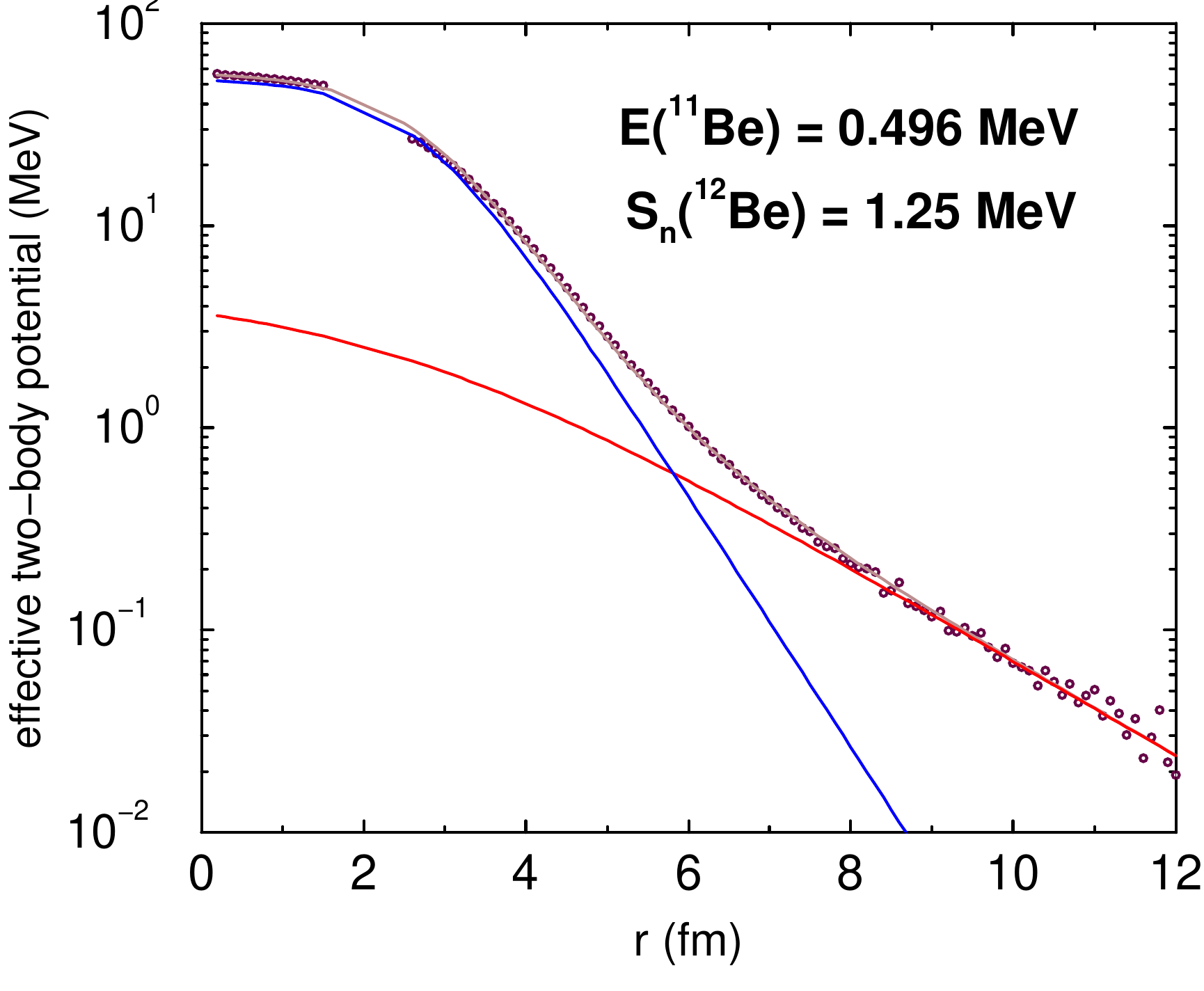}
\caption{\label{label}Effective local $n-^{11}$Be potentials
for pseudo $^{12}$Be fitted by a sum of two Woods-Saxon potentials.}
\end{minipage} 
\end{figure}

\section{Three-body calculations}

We have performed three-body calculations for $^{12}$Be ($^{10}$Be+n+n), 
$^9$C ($^7$Be+p+p) and $^{18}$Ne  ($^{16}$O+p+p) 
assuming no structure for the
$^{10}$Be, $^7$Be and $^{16}$O cores but taking
antisymmetrization  into account
using the Pauli Projection technique \cite{PP}. The three-body 
Schr\"odinger equation has been solved by expanding the 
wave functions  onto a hyperspherical harmonics basis.
The overlap integrals $\la^{11}$Be$|^{12}$Be$\ra$, 
$\la^{8}$B$|^{9}$C$\ra$
and $\la^{17}$F$|^{18}$Ne$\ra$ have been obtained by overlapping
the three-body and two-body wave functions. The $N$-core Woods-Saxon
potentials
have been fitted to chosen separation energies and the GPT
potential \cite{GPT} has been used for the
$NN$ interaction. The Coulomb $p$-core and $pp$
potentials have been also included
for $^{9}$C and $^{18}$Ne.  

{\bf Pseudo $^{12}$Be}. First of all the convergence of the overlap
in the asymptotic  region has been tested for a simplified case 
of $^{12}$Be in which the deformation and excitation of the
$^{10}$Be core have been neglected and the $n-^{11}$Be interaction was 
 zero in
all partial waves except for $\ell = 0$. This has enabled us
to include terms up to $K_{\max} = 120$   in the hyperspherical expansion
and to obtain converged values for $\la^{11}$Be$|^{12}$Be$\ra$ 
with high accuracy up to 30 fm.
The  $n-^{10}$Be  potential   was fitted  
either to the experimental value of
$S_{1N}(^{11}$Be), 496 keV, or to a much smaller
value of 102 keV to check whether the effect from the
$nn$ correlations increases with weaker binding. 
In this particular case,
the $nn$ potential was represented by just one Gaussian that
gives a scattering length of $-20$ fm. The three-body energies obtained for
$^{12}$Be are shown in Table 1. The overlap integrals 
$\la^{11}$Be$|^{12}$Be$\ra$, renormalized to unity,
 are plotted as ratios to their asymptotics
in Fig.1a where they are compared to those calculated
in a potential model with standard geometry $r_0$ = 1.25 fm
and $a$ = 0.65 fm.
Also, the r.m.s. radii of these overlaps are given in Table 1 in comparison
with standard values.
The abnormal behaviour is clearly seen for both three-body calculations 
and, as expected, the abnormalities are stronger
for a smaller $S_{1N}(^{11}$Be). Fig.2 shows effective local potentials 
 (open circles) obtained by inversion of the two-body
Schr\"odinger equation. 
They can be  fitted
by a sum of two  Woods-Saxon potentials one of which has unusually 
large diffuseness (see Table 2).

{\bf Realistic $^{12}$Be}. To take  into account the deformation
and the excitation of the $^{10}$Be core,
we have repeated the calculations from \cite{Nun02} for the case EXC1.
The model space have been increased up to $K_{\max}$ = 34 which has provided
a reasonable stability for the overlap $\la^{11}$Be$|^{12}$Be$\ra$
at   $5 \leq r \leq 10$ fm. 
The abnormal surface behaviour in $\la^{11}$Be$|^{12}$Be$\ra$
is seen for all three states of the   $^{11}$Be residue
(see Fig. 1b) and  the
effective local $n -  ^{11}$Be potentials, the parameters of
which are shown in Table 2, have
 long-range terms  although their diffuseness are smaller than 
in the pseudo $^{12}$Be case. 
The   abnormalities arise due to the correlations between the
two neutrons which are visualised in the plot of the wave function
densities
in Fig. 3. When the c.m. of two neutrons is far away from
the $^{10}$Be core, the probability that these neutrons
are close to each other is high.

\begin{figure}[h]
\begin{minipage}{40pc}
\includegraphics[width=36pc]{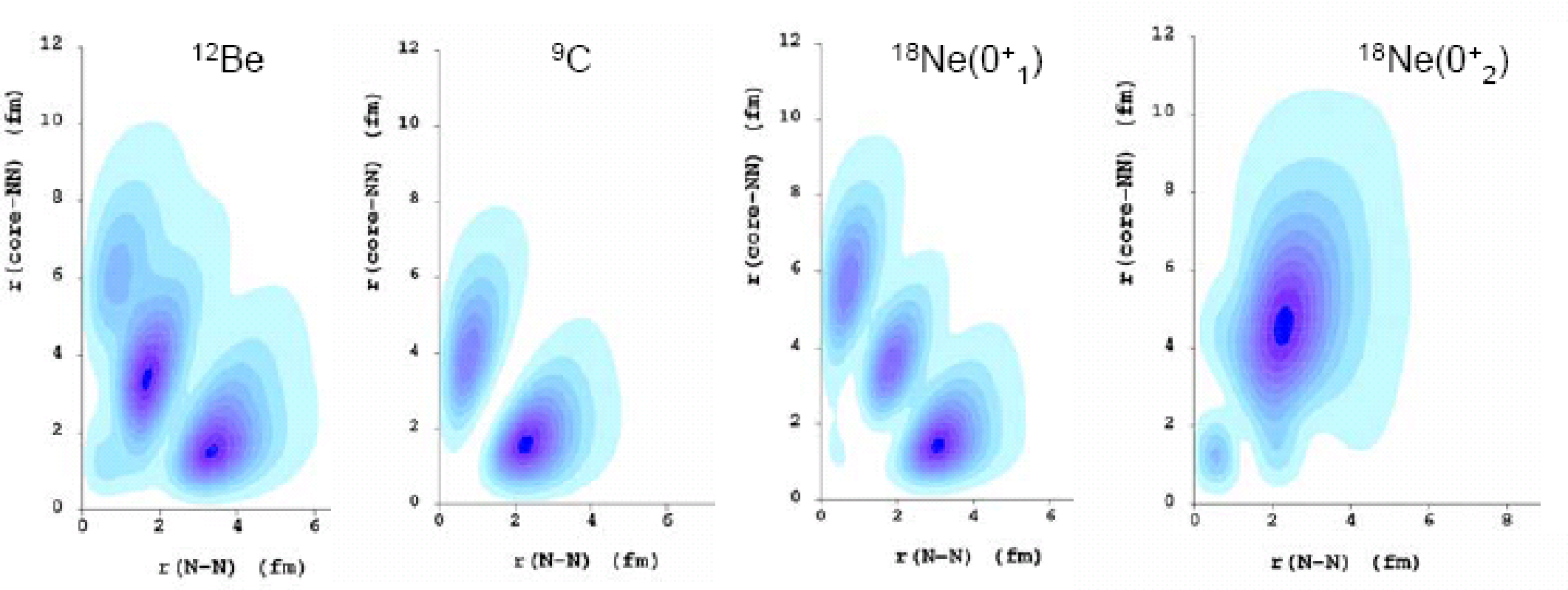}
\caption{\label{label}Probability densities of the wave functions
obtained in three-body calculations for realistic
$^{12}$Be, $^{18}$Ne and
pseudo $^9$C.}
\end{minipage}\hspace{1pc}%
\end{figure}

\begin{table}[b]
\caption{\label{ex}One- and two-nucleon separation energies,
$S_{1N}$ an $S_{2N}$ (in MeV),
and r.m.s. radii (in fm)
for overlaps $\la A-1 | A \ra$ calculated in a three-body model.  
R.m.s. radii $\la r^2\ra^{1/2}_{st}$ from the fixed standard
Woods-Saxon potential are also given.}
\begin{center}
\begin{tabular}{lllllll}
\br
nucleus $A$ & nucleus $A$$-$$1$ &
$S_{1N}(A$$-$$1)$  & $S_{1N}(A)$ & $S_{2N}(A)$ & $\la r^2\ra^{1/2}$
& $\la r^2\ra^{1/2}_{st}$ \\
\mr
pseudo $^{12}$Be & $^{11}$Be($\frac{1}{2}^+$) & 
                   0.102 & 0.56 & 0.656 & 7.650 & 6.983\\
pseudo $^{12}$Be & $^{11}$Be($\frac{1}{2}^+$) & 
                   0.496 & 1.25 & 1.76 & 5.807 & 5.423 \\                   
real $^{12}$Be & $^{11}$Be($\frac{1}{2}^+$) & 
                    0.500  & 3.141 &  3.641 & 4.412 & 4.298 \\
               & $^{11}$Be($\frac{1}{2}^-$) & 
                    0.180  & 3.461 &  3.641 & 3.588 & 3.473 \\
               & $^{11}$Be($\frac{5}{2}^+$) & 
                   $-$1.09   & 4.731 &  3.641 & 3.738 & 3.244 \\
pseudo $^{9}$C & $^{8}$B & 
                   0.149 & 4.526 & 4.675 & 3.175 & 3.016 \\ 
 $^{18}$Ne($0^+_1$) & $^{17}$F($\frac{5}{2}^+$) & 
                     0.603 & 2.974 &   3.577 & 3.602 & 3.337 \\
   & $^{17}$F($\frac{1}{2}^+$) & 
                     0.176 & 3.401 &   3.577 & 4.068 & 3.602 \\
   $^{18}$Ne($0^+_2$) & $^{17}$F($\frac{5}{2}^+$) & 
                     0.603 & 2.974 &   1.141 & 3.511 & 3.426 \\
   & $^{17}$F($\frac{1}{2}^+$) & 
                     0.176 & 3.401 &   1.141 & 4.486 & 3.947 \\
\br
\end{tabular}
\end{center}
\end{table}

{\bf Pseudo $^9$C}. Our attempts to use the proper spin 
$J^{\pi} = \frac{3}{2}^-$for the $^{7}$Be
core have been successful only for  small model spaces that cannot
provide convergence for the overlap in the region 
of interest. Therefore, we have used $J^{\pi} =0^+$ for $^{7}$Be
and neglected the $p$-$^{7}$Be spin-orbit interaction.
This overbinds $^9$C but
 enables us to increase the model space up to $K_{\max}$ = 50, 
 thus providing
high accuracy for $\la^{8}$B$|^{9}$C$\ra$ up to 20 fm.
The overlap converges 
to its asymptotic form slower than
in the standard case but not as slow as for $^{12}$Be. The two
protons stay close to each other when both are far away from the
$^7$Be core but the weight of this configuration is smaller
than in $^{12}$Be (see Fig.3).

{\bf $^{18}$Ne}. The $^{18}$Ne
wave functions have been calculated for the ground
and first excited $0^+$ states in a model space with $K_{\max}$ = 50.
The $pp$ correlations outside the nuclear interior are clearly seen
in the $^{18}$Ne wave function and the overlaps differ from the
standard ones in the surface region. However, this difference is much
smaller than for $^{12}$Be, despite the long-range character of the Coulomb
force. The  three-body dynamics results mainly in  an increase of the
overlap r.m.s. radius, which is stronger in the $s$-wave
where the centrifugal barrier is absent.

\begin{table}[t]
\caption{\label{ex}The depth $V_i$ (in MeV), radii $r_i$ and
diffusenesses $a_i$ (in fm) of the local effective potentials for
the overlap $\la A-1 | A \ra$ parametrised
as a sum of two Woods-Saxon central potentials, $V_0(r)+V_1(r)$,
and a spin-orbit potential $V_{so}(r)$.
}
\begin{center}
\begin{tabular}{lllllllllll}
\br
 nucleus $A$ &  $A$$-$$1$ &
  $V_0$ & $r_0$ & $a_0$ & $V_1$ & $r_1$ & $a_1$
 & $V_{so}$ & $r_{so}$ & $a_{so}$\\
\mr
pseudo $^{12}$Be & $^{11}$Be($\frac{1}{2}^+$) &     53.5 & 1.20 
& 0.70 & 4.8 & 0.989 & 1.85 \\
pseudo $^{12}$Be & $^{11}$Be($\frac{1}{2}^+$) &     55.9 & 1.07 
& 0.82 & 1.5 & 1.124 & 3.1 \\
real $^{12}$Be & $^{11}$Be($\frac{1}{2}^+$) &     50.0 & 1.33 
& 0.75 & 6.9 & 1.17 & 1.4 \\
& $^{11}$Be($\frac{1}{2}^-$) &   50.0 & 0.99 
& 0.79 & 9.8 & 1.12 & 1.55 & 8.85 & 1.08 & 0.80 \\
pseudo $^{9}$C & $^{8}$B &  54.24 & 1.10 & 0.81 & 2.7 & 1.2 & 1.5
& 8.91 & 1.10 & 0.81 \\
\br
\end{tabular}
\end{center}
\end{table}

\section{Non-standard overlaps in nuclear reactions}

\begin{figure}[b]
\includegraphics[width=14pc]{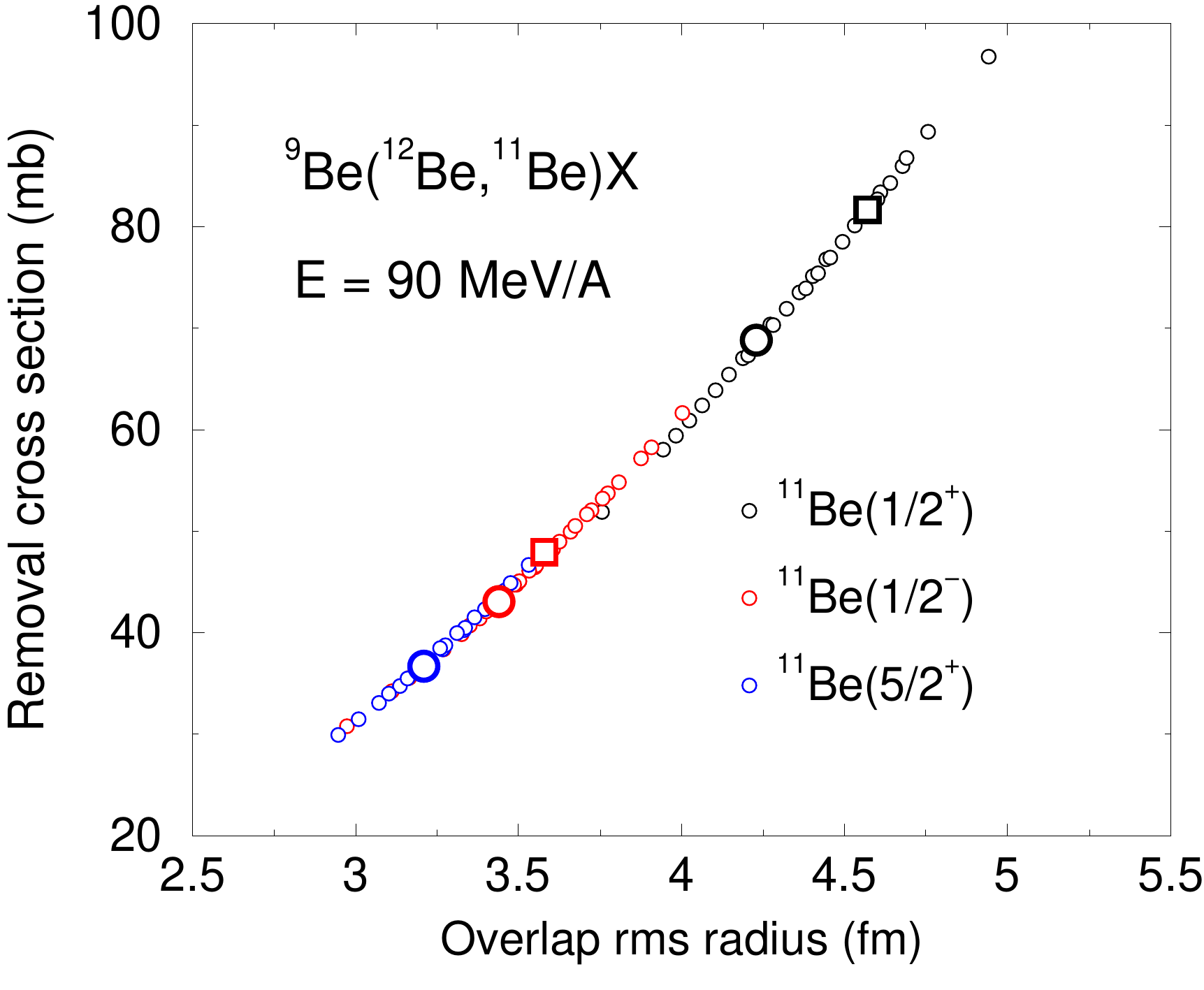}\hspace{2pc}%
\begin{minipage}[b]{22pc}\caption{\label{label}
The knockout cross sections populatingthe final states $^{11}$Be($\frac{1}{2}^+$) (in black), $^{11}$Be($\frac{1}{2}^-$)
(in red) and $^{11}$Be($\frac{5}{2}^+$) (in blue) calculated with  two-body $n-^{11}$Be wave functions calculated for  a range of different
geometries of the Woods-Saxon potentials (circles). The cross sections corresponding to the standard Woods-Saxon geometry are shown by open circles while those calculated with a sum of two Woods-Saxon potentials from Table 2 are shown by open squares.}
\end{minipage}
\end{figure}

{\bf Nucleon knockout reactions}  
have recently become a popular tool
to determine spectroscopic factors. We have 
calculated  the cross sections and the parallel momentum
distributions for   $^9$Be($^{12}$Be,$^{11}$Be)X  
at 90 MeV/nucleon using the $\la^{11}$Be$|^{12}$Be$\ra$ overlap
obtained in three-body calculations.  
The  parallel momentum distributions from the 
three-body $\la^{11}$Be$|^{12}$Be$\ra$ overlap agree with those from the
Woods-Saxon potential, provided they have the same r.m.s. radius. The
slow convergence of $\la^{11}$Be$|^{12}$Be$\ra$
to its asymptotic form cannot be seen in such an
experiment. The momentum distributions   depend on the 
radius of the overlap 
and thus can be used to determine this
if measured with high precision. The cross
sections scale almost linearly with the r.m.s. radius of the overlap
(see Fig.4) and,  
therefore, at these energies the effect from surface
abnormalities in the $\la^{11}$Be$|^{12}$Be$\ra$ overlap manifests
itself only as a change in the 
overall norm of the calculated cross sections. This will therefore
influence   spectroscopic factors deduced from comparisons
between calculated and measured cross sections.
 
{\bf Transfer reactions} 
are sensitive to the surface part of the overlap 
and thus could be a good tool to study 
non-standard single-particle motion in the core+2N cluster systems. 
A subclass of such
reactions, peripheral
transfer reactions, 
determine asymptotic normalization coefficients in the overlap tails
thus constraining  the overlap r.m.s. radius.  

{\bf Low-energy direct radiative capture}. Proton capture
cross sections are often sensitive only to the asymptotic
region  of the one-nucleon overlap and thus possible abnormal 
surface behaviour will not influence energy dependence of such
reactions. However, 
the contribution of the surface region to the neutron
capture amplitude can be  more important.
We have performed two-body calculations of the
$^{11}$Be(n,$\gamma$)$^{12}$Be  cross sections 
using three
different  $n-^{11}$Be interactions:
(i) the standard Wood-Saxon potential;
(ii) the effective local potential from Table 2 presented by two 
Woods-Saxon potentials
and  (iii) a single Woods-Saxon potential that gives exactly the
same r.m.s. radius for the $n-^{11}$Be wave function
as the effective local one. Both the absolute values 
and the energy dependence of
the cross sections, shown in Fig. 5, 
are very sensitive to the shape of the overlap.

\begin{figure}[t]
\includegraphics[width=12pc]{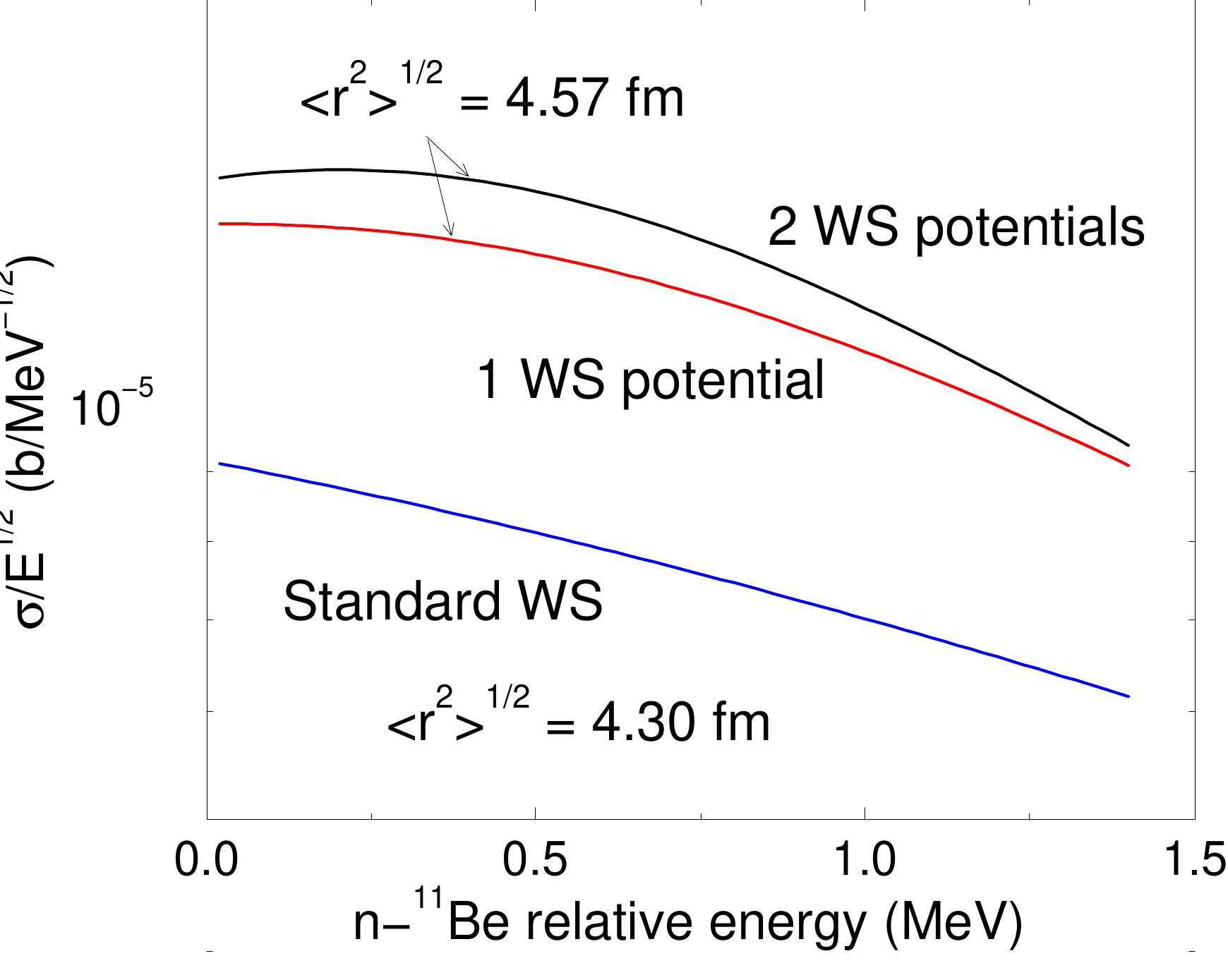}\hspace{2pc}%
\begin{minipage}[b]{18pc}\caption{\label{label}
The  cross sections for the $^{11}$Be(n,$\gamma$)$^{12}$Be reaction 
divided by square root of the $n-^{11}$Be energy calculated with three different  $n-^{11}$Be potentials. The $\sigma/E^{1/2}$ is shown in logarithmic scale. }
\end{minipage}
\end{figure}

\section{Challenges for theory and experiment}

The exact dynamical three-body calculations have confirmed the
possibility of slower radial convergence of the one-nucleon overlaps
to their asymptotic form in  core+2N cluster systems due
to the strong correlations between the two valence nucleons. 
This phenomenon has 
the strongest effect for final core+N  systems   
in relative $l=0$  states and when the Coulomb interaction is absent.
The
non-standard behaviour has consequences for the 
determination of spectroscopic
factors from nucleon removal reactions and predictions of 
neutron capture rates on weakly-bound $s$-wave nuclei at
stellar energies. Therefore, theory must  make reliable 
predictions for non-standard overlaps. 

The main challenge for theory is to treat explicitly the three-body  
dynamics within a many-body  object.  However, even when internal
structure of the core is neglected, the model space, needed to
describe the nucleon motion at large distances in coupled-channel
hyperspherical
calculations, becomes  huge. Convergence accelerating methods
should be developed  to describe properly the  core
deformation and excitations within the three-body model. 
Unsolved is the question of how   antisymmetrisation  
influences non-standard behaviour of the overlap. The widely used 
phenomenological
shell model does not generate  single-particle wave functions at all
so no abnormalities can be seen there.  
No  mean field based theory can reproduce this effect. 
On the other hand, ab-initio approaches do not yet have 
sufficient accuracy to study such effects at large distances.
For example, keeping in mind that to see unambiguously 
abnormalities at $5 \leq r \leq 10$ fm in three-body calculations 
 the $K_{\max}$ should be at least 30, means  that within  models
of the no-core shell model type  at least 30 additional major 
shells are needed which do not influence the total binding energy.
This does not appear feasible at the moment.
Three-body dynamics can be  included explicitly in microscopic
cluster models by hand 
\cite{pdesc}. However, at present, such models can cope only
with the simplest oscillator internal core structure
and very simple NN interactions.   

Observation of predicted three-body
effects in one-nucleon overlaps
can be a very difficult task because such effects can be 
obscured by unsufficient knowledge of reaction mechanisms
and uncertainties of other inputs to reaction amplitudes. However,
unambigious experimental confirmation  either of their
presence or absence is
very important for our understanding of nuclear dynamics
in near-drip line nuclei.

This work was performed under the  UK grants
EP/C520521/1 and EP/E036627/1 and in
the Lawrence Livermore National Laboratory 
under DoE Contract DE-AC52-07NA27344.

\section*{References}



\begin{thebibliography}{3}
\bibitem{Tim03} Timofeyuk N K,  Blokhintsev L D and  Tostevin J A
2003 {\it Phys. Rev. C} {\bf 68} 021601(R)
 \bibitem{Blo01} Blokhintsev L D 2001 
{\it Bull. Russ. Acad. Sci. Phys}. {\bf 65} 77 
\bibitem{PP} Thompson I.J et al 2000 {\it Phys. Rev. C} {\bf 61} 24318
\bibitem{GPT} Gogny D, Pires P and De Tourreil R 1970
{\it Phys. Lett.} {\bf 32} 571
\bibitem{Nun02} Nunes F M,   Thompson I J and   Tostevin J A 2002
  {\it Nucl. Phys. A} {\bf 703} 593
\bibitem{pdesc} Korennov S and Descouvemont P 2004
{\it Nucl. Phys.}  A 740  249  
\end{thebibliography}
\end{document}